\documentclass{jkas}

\def\year{2018} 
\def\month{April} 
\def\volume{51} 
\def\issue{2} 
\def\beginpage{37} 
\def\endpage{48} 
\def\received{March 12, 2018} 
\def\accepted{April 10, 2018} 
\date{Received \received; accepted \accepted}

\def\kms{~{\rm km~s^{-1}}}
\def\ergs{~{\rm erg~s^{-1}}}
\def\cm3{~{\rm cm^{-3}}}

\def\smass{{M_{\odot}}}
\def\Mdot{\dot {M}}
\def\vterm{v_{\infty}}
\def\vesc{v_{\rm esc}}
\def\Msunyr{\rm M_{\odot}~yr^{-1}}
\def\Mcl{M_{\rm cl}}

\usepackage{flushend}

\title{The Contribution of Stellar Winds to Cosmic Ray Production}

\author[1]{Jeongbhin Seo}
\author[1]{Hyesung Kang}
\author[2]{Dongsu Ryu}

\affil[1]{Department of Earth Sciences, Pusan National University, 2  Busandaehak-ro, Geumjeong-gu, Busan 46241, Korea; \email{hskang@pusan.ac.kr}}
\affil[2]{Department of Physics, School of Natural Sciences, UNIST, 50 UNIST-gil, Ulsan 44919, Korea; \email{ryu@sirius.unist.ac.kr}}

\def\corrauthor{H. Kang}

\def\runningauthor{Seo et al.}

\def\runningtitle{Stellar Winds from Massive Stars}

\def\keywords{stars: massive --- stars: mass-loss --- stars: winds --- acceleration of particles --- shock waves --- cosmic rays }

\def\abstracttext{
Massive stars blow powerful stellar winds throughout their evolutionary stages from the main sequence to Wolf-Rayet phases.
The amount of mechanical energy deposited in the interstellar medium by the wind from a massive star can be comparable to the explosion
energy of a core-collapse supernova that detonates at the end of its life.
In this study, we estimate the kinetic energy deposition by massive stars in our Galaxy
by considering the integrated Galactic initial mass function and modeling the stellar wind luminosity.
The mass loss rate and terminal velocity of stellar winds during the main sequence, red supergiant, and Wolf-Rayet stages
are estimated by adopting theoretical calculations and observational data published in the literature.
We find that the total stellar wind luminosity due to all massive stars in the Galaxy is about ${\mathcal L}_{\rm w} \approx 1.1\times 10^{41} \ergs$,
which is about 1/4 of the power of supernova explosions, ${\mathcal L}_{\rm SN} \approx 4.8\times 10^{41} \ergs$.
If we assume that $\sim 1-10$ \% of the wind luminosity could be converted to Galactic cosmic rays (GCRs)
through collisonless shocks such as termination shocks in stellar bubbles and superbubbles, colliding-wind shocks in binaries,
and bow-shocks of massive runaway stars,
stellar winds might be expected to make a significant contribution to GCR production, though lower than that of supernova remnants.
}

\begin{document}
\jkashead 

\section{Introduction}
\label{intro}

Massive stars with $M_{\rm ZAMS}\gtrsim 10\smass$ deposit a significant amount of mechanical energy into the interstellar medium (ISM)
through stellar winds during main sequence (MS), red supergiant (RSG), and Wolf-Rayet (WR) stages,
as well as supernova (SN) explosions at the end of their lives \citep[e.g.,][]{ekstrom2012,georgy2012,smith2014,yoon2015}.
Hereafter, $M_{\rm ZAMS}$ refers to the stellar mass at zero-age MS (ZAMS).
At each stage of their lives, the mechanical power of stellar winds due to a massive star can be characterized by the mass loss rate, $\Mdot$,
and the terminal velocity, $\vterm$.
For a star with $35\smass$, for instance,
(1) $\Mdot \sim 10^{-6}~\Msunyr$ and $\vterm\sim 3,000\kms$ during the MS stage,
(2) $\Mdot \sim 10^{-4}~\Msunyr$ and $\vterm\sim 25\kms$ during the RSG stage,
(3) $\Mdot \sim 3\times 10^{-5}~\Msunyr$ and $\vterm\sim 5,000\kms$ during the WR stage \citep{garcia1996b,smith2014}.
The so-called {\it wind luminosity}, $L_{\rm w}\equiv (1/2)\Mdot \vterm^2$, ranges $\sim 10^{34} - 10^{38} \ergs$ during the MS stage
for $M_{\rm ZAMS}\approx 15-120~\smass$ \citep{georgy2013}.
The uncertainties on mass loss, along with convective overshooting, rotation, and
magnetic fields, greatly hinder full understanding of the evolution of massive stars
  \citep[e.g.,][]{yoon2010,ekstrom2012, georgy2012}.
For example, depending on their rotation,
stars with $\gtrsim 20-25~\smass$ are expected to experience the WR phase,
but those with $\gtrsim 32-40~\smass$ enter the WR phase without going through the RSG phase \citep{georgy2012}.

The interaction of stellar winds with the circumstellar medium has been studied extensively by hydrodynamic simulations
that are equipped with the mass loss parameters (i.e., $\Mdot$ and $\vterm$) from stellar evolution calculations \citep[e.g.,][]{garcia1996a, garcia1996b, freyer2003,georgy2013}.
The idealized spherical structure of the so-called {\it stellar wind bubble} can be found in many previous studies such as \citet{weaver1977} and \citet{freyer2003}.
Typically, the expanding wind is terminated by a (reverse) termination shock that propagates into the wind flow, while a forward shock expands into the photo-ionized circumstellar medium.
The termination shock decelerates and heats the wind gas to $10^6-10^8$~K, creating a hot bubble around the central star.
If we assume that an average of $\langle L_{\rm w}\rangle \sim 10^{36} \ergs$ per star is deposited to the ISM through stellar winds
and that there are $10^5$ OB stars in the Galaxy, then the total wind power from massive stars
is roughly in the order of ${\mathcal L}_{\rm w} \sim 10^{41} \ergs$.
On the other hand, with the SN explosion rate (SNER) of a few per century \citep{Reed05} and the explosion energy of $E_{\rm SN}\sim10^{51}$~ergs,
the energy deposition rate of SNe in the Galaxy is estimated as ${\mathcal L}_{\rm SN} \sim 3\times 10^{41} \ergs$.
So the contribution of the mechanical energy from stellar winds could be comparable to that from SN explosions in the Milky Way.

Galactic cosmic rays (GCRs) with energies lower than $\sim 10^{6}$ GeV/nucleon are thought to be accelerated mainly within the Galactic disk \citep[see][for reviews]{blandford1987,hillas2005, drury2012}.
They are transported into the Galactic halo and escape from our Galaxy through galactic winds with a roughly constant leakage rate.
From the observed ratios of the secondary to primary CRs and the observed anisotropy of CR distribution,
the power of CR sources required to maintain the energy density of GCRs is estimated to be $3-8\times 10^{40} \ergs$ \citep{strong2010, drury2012}.
Dominant source candidates that can replenish such escape of GCRs are all related to massive stars:
supernova remnants (SNRs), stellar winds, and pulsar winds.

Nonthermal particles are known to be accelerated via diffusive shock acceleration (DSA) at collisonless shocks
that are ubiquitous in astrophysical environments, from the Earth's bow-shock to merger shocks in galaxies clusters \citep{drury1983}.
The DSA efficiency at such shocks depends mainly on the shock Mach number, magnetic field obliquity angle, and strength of MHD turbulence
responsible for particle scattering \citep[e.g.,][]{treumann2009}.
For instance, CR protons are accelerated efficiently at quasi-parallel shocks, while CR electrons are accelerated preferentially at quasi-perpendicular shocks \citep{riqu11}.
Recent plasma simulations have indicated that about 10\% of the shock kinetic energy is transferred to the energy of CR protons at strong shocks
with the shock Mach number $M_s\gtrsim10$ and `quasi-parallel' magnetic field configuration \citep{caprioli14}.

It is well established that CR ions with atomic charge $Z_{\rm a}$ can be accelerated up to the knee energy of $3\times 10^{15}Z_{\rm a}$~eV by strong SNR blast
shocks and that approximately 10 \% of the explosion energy can be transferred to GCRs in the Galaxy \citep[see][for recent reviews]{blasi2013,caprioli2015}.
Similarly, a significant fraction of wind mechanical power, $L_{\rm w}$, from massive stars in pre-supernova stages
is expect to be converted to GCR energy through DSA at
termination shocks inside stellar wind bubbles \citep{casse1980,volk1982,drury1983},
shocks in particle-accelerating colliding-wind binaries (PACWBs) \citep{debecker2017},
and bow-shocks of massive runaway stars \citep{delvalle2012}.
CR electron acceleration at strong shocks produced by colliding winds in binary (or multiple) stellar systems have been observed \citep{debecker2013}.
In addition, bow-shocks produced by the interaction between strong winds from massive runaway stars and the ISM may provide a minor contribution to GCR production as well \citep{delvalle2015}.
On the other hand, pulsar winds are ultra-relativistic plasmas composed of primarily electron-positron pairs,
and termination shocks are predominantly {\it perpendicular} with toroidal magnetic fields.
Thus, mainly CR leptonic components are thought to be accelerated via shock drift acceleration and/or magnetic reconnection
at pulsar wind termination shocks \citep{Amato2014,Sironi2017}.


In stellar wind bubbles around massive stars, typical termination shocks with $\vterm \sim 3\times 10^3 \kms$ are strong shocks that stop
the wind flow with $\rho_{\rm w}\propto r^{-2}$ and $T_{\rm w}\sim 10^4$~K, photo-ionized by the central star \citep{weaver1977}.
Yet, contrary to SNRs, the conversion of wind energy to CRs at termination shocks has not been estimated quantitatively.
Unlike SNR blast waves with their relatively well preserved spherical symmetry, the structures inside wind bubbles are complex 
with the termination shock, contact discontinuity, and multiple shells due to winds at different stages. 
They are prone to several instabilities such as Rayleigh-Taylor, thin-shell, and thermal instabilities \citep{garcia1996b}.
Moreover, the wind flow may contain strong MHD turbulence and density clumps.
So it is difficult to estimate the key physical parameters for the DSA process such as the radius, lifetime, magnetic field strength and
obliquity of the termination shock as well as MHD turbulence of the preshock wind plasma.
Moreover, if the star moves relative to the ISM \citep{meyer2014}, or if the circumstellar medium is not uniform,
then the simple spherical geometry of the wind bubble is distorted.
Thus, it is very challenging to calculate the overall DSA efficiency at such complex and turbulent structures.
As far as we know, an accurate
estimation for the CR conversion rate of wind mechanical energy at the termination shock is not available in the literature.

Another complication is the fact that massive stars form not in isolation, but as binary systems, OB associations, or stellar clusters
inside dense molecular clouds, which may lead to interacting multiple winds or superbubbles \citep[e.g.,][]{debecker2007, zinnecker2007, vanmarlet2012}.
The DSA efficiency at shocks formed in the wind-wind interaction region of PACWBs is likely to be higher than that at termination shocks around individual member stars.
However, it depends on the fraction of the shock surface area of the wind-wind interaction region, in addition to the uncertain DSA parameters
such as magnetic field obliquity and preshock turbulence \citep{debecker2013}.
Note that PACWBs have been confirmed observationally by nonthermal synchrotron radiation,
although direct observational evidence for nonthermal emission from individual wind bubbles has yet to be established.

The bulk of core-collapse SNe are observed to be clustered in superbubbles, which are created by previous episodes of stellar winds and
SN explosions inside OB associations \citep{higdon2013}.
It has been shown that the observed composition of GCRs can be explained by the CR acceleration at SNRs expanding inside metal-enriched
superbubbles \citep[e.g.,][]{higdon1998,Binns05,bykov2014}.
In addition, gamma-ray emission due to pion production in pp collisions in the Cygnus cocoon, as detected by the Fermi-LAT telescope,
is interpreted as the first direct evidence for the CR acceleration in a superbubble with OB-star complexes  \citep{ackermann2011}.
Obviously, it would be very difficult to quantify the CR acceleration by shocks associated with stellar winds, separately from SNRs,
inside these complex superbubbles. Hence, these issues are not addressed here.

Considering these issues, here we assume that approximately $1-10$\% of the wind luminosity could be transferred to GCRs
at various shocks associated with massive stars including termination shocks inside stellar bubbles and superbubbles,
shocks formed by colliding winds in multiple star systems,
and bow-shocks around runaway stars.

In this study, we attempt to estimate the relative importance of wind mechanical energy deposition at different stages,
i.e., MS, RSG, and WR phases.
To that end, we first model the wind luminosity at different stages as a function of stellar mass by adopting theoretical estimates or observational data
for $\Mdot$ and $\vterm$.
Adopting the galaxy-wide initial mass function of massive stars, we then estimate the number of massive stars existing in the Galactic disk
and their kinetic energy contribution due to stellar winds as a function of stellar mass.
Finally, we compare the total wind power deposited from stellar winds to the SNR explosion power in the Milky Way.

In the next section, we describe how we model the integrated Galactic initial mass function, mass loss rate, terminal velocity, and luminosity of stellar winds at different stellar types.
In Section \ref{results}, we calculate the integrated wind power due to all massive stars in the Galaxy.
Section \ref{summary} presents a brief summary.

\begin{figure*}[t!]
\centering
\includegraphics[trim=2mm 2mm 2mm 2mm, clip, width=160mm]{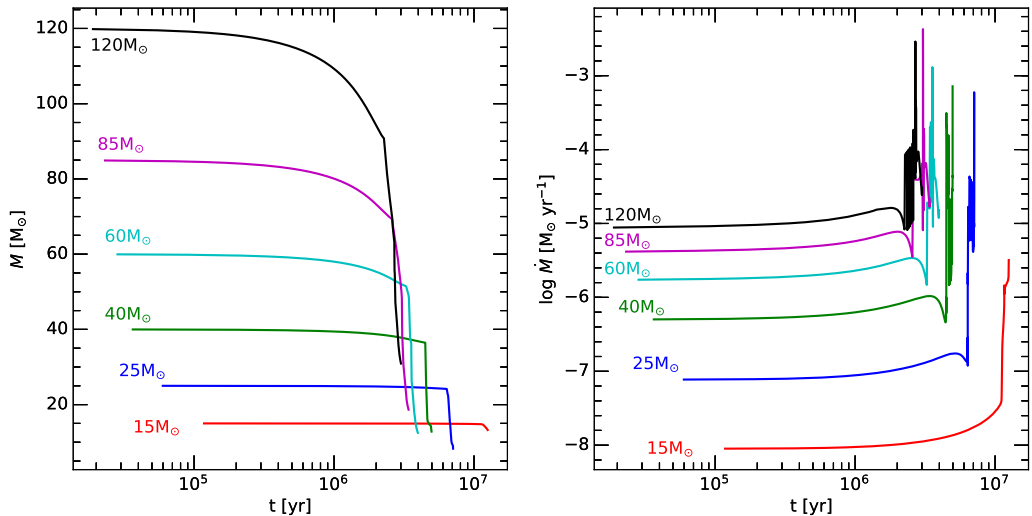}
\caption{Time evolution of the stellar mass $M(t)$ (left) and the mass loss rate $\Mdot(t)$ (right)
taken from the grid of the stellar evolution computation for nonrotating stars presented by \citet{ekstrom2012}.
Here $M(t)$, $\Mdot$, and $t$ are expressed in units of $\smass$, $\Msunyr$, and years, respectively.
\label{fig:f1}}
\end{figure*}

\section{Models}
\label{models}

In this section, we define the mass distribution function of massive stars in the Galaxy. Then
we explain how we model the mass loss rate and the terminal velocity of stellar winds at different stages.
We use them to estimate wind mechanical energy deposition as a function of stellar mass.

\subsection{Integrated Galactic Initial Mass Function}
\label{igimf}

The initial mass function (IMF), $\xi(m)=dn/dm$, describes the observed distribution of the initial mass of stars in groups such as stellar clusters
\citep[see][for a review]{Kroupa13}.
Here $\xi(m) dm$ is the number of stars in the unit volume whose initial mass, $m\equiv M_{\rm ZAMS}$, lies between $m$ and $m+dm$.
It is very difficult to predict the IMF theoretically, since star formation involves the complex interplay of many
physical processes including gravity, hydrodynamics, radiative transfer, turbulence, magnetic fields, and external radiation field
\citep[see][for a review]{McKee2007}.
However, the observed IMF is found to be remarkably universal in a wide range of environments, and can be represented well by the following canonical power-law form,
\begin{equation}
\xi(m)=A\cdot m^{-\alpha},
\label{imf}
\end{equation}
where the power-law index is $\alpha\approx 2.3-2.7$ for stars with $1~\smass \le m\le 150~\smass$
\citep{Salpeter55,Schmidt59,Miller79,Scalo86,Kroupa02}.
Hereafter, $m$ is expressed in units of $\smass$,
so the normalization factor $A$ is given in units of ${\rm pc}^{-3}$.

Massive stars form predominantly inside OB associations or stellar clusters.
The observed distribution function of cluster mass, $\Mcl$, formed in our Galaxy can also be described by a similar power-law function,
\begin{equation}
\xi_{\rm cl}(\Mcl)=A_{\rm cl}\cdot \Mcl^{-\beta},
\label{Mclmf}
\end{equation}
for the cluster mass range of $ M_{\rm cl,min}\le M_{\rm cl}\le M_{\rm cl,max}$ \citep{Kroupa13}.
According to \citet{Weidner13},
the power-law index is $\beta \approx 2.0$ for the star formation rate (SFR) of $\sim 1~\Msunyr$,
and the maximum cluster mass also depends on SFR as
\begin{equation}
M_{\rm cl,max}(\smass)=8.5\times10^4\cdot \left( \frac{{\rm SFR}(t)}{\Msunyr} \right)^{0.75}.
\label{Mcl}
\end{equation}
The smallest mass of observed clusters can be taken as $ M_{\rm cl,min} \approx 5~\smass$.

Following \citet{Kroupa13}, we define the Integrated Galactic IMF (IGIMF) as the galaxy-wide IMF at a given time for
all stars contained in the entire population of stellar clusters in the Galaxy:
\begin{eqnarray}
& &\xi_{\rm IGIMF}(m;t)=\nonumber\\
& &\int_{\rm M_{cl,min}}^{M_{\rm cl,max}(t)} \xi(m<m_{\rm max}(\Mcl))\cdot\xi_{cl}(\Mcl)d\Mcl.
\label{igimfeq}
\end{eqnarray}
Here, $m_{\rm max}(\Mcl)$ is the maximum mass of the member stars contained in a cluster with $\Mcl$.
In general, the upper limit of the integration, $M_{\rm cl,max}(t)$, depends on time, for instance, as given in Equation (\ref{Mcl}),
since SFR changes with time.
So depending on the time variation of SFR and ${M_{\rm cl,max}(t)}$ in our Galaxy,
IGIMF can have a power-law distribution steeper than the canonical IMF \citep{Weidner13}.

The determination of the normalization factors, $A$ and $A_{\rm cl}$, as well as the power-law indices, $\alpha$ and $\beta$,
based on observed stellar populations is limited due to severe interstellar extinction,
since massive stars are born in the Galactic disk and located mainly near spiral arms.
However, those parameters can be estimated indirectly by comparing certain theoretical predictions with observed quantities.
For example, the chemical composition of the ISM is a product of the sum of all star-formation events
and ensuing chemical enrichment throughout the history of our Galaxy \citep{Kroupa13}.

Here, adopting the results of \citet{Weidner13}, IGIMF is assumed to have the power-law form of
$\xi_{\rm IGIMF}(m)\propto m^{-2.6}$ for a SFR of $1~\Msunyr$.
Thus we assume that the distribution function of all massive stars contained in the Galactic disk at the present time
has the following form:
\begin{equation}
N(m)=A_{\rm OB}\cdot m^{-2.6}.
\label{pdimf}
\end{equation}
So $N(m)dm$ represents the total number of stars with the initial mass between $m$ and $m+dm$ in the present-day Galaxy.
As defined above, $m=M_{\rm ZAMS}$ is expressed in units of solar masses, so the normalization factor,
$A_{\rm OB}$, is dimensionless.

We can estimate $A_{\rm OB}$ approximately, using the fact that SNER in our Galaxy is 1-2 in 100 years.
In other words, if each star heavier than $10~\smass$ explodes as a core-collapse SN after its MS lifetime, $\tau_{\rm MS}(m)$,
SNER can be calculated approximately by
\begin{equation}
{\rm SNER} \approx\int_{10{\rm M_{\odot}}}^{150{\rm M_{\odot}}} \frac{N(m)}{\tau_{\rm MS}(m)} dm.
\label{sner}
\end{equation}
For the MS lifetimes we adopt the results of the stellar evolution calculation due to \citet{Schaller92}, 
modelled by the following fitting form \citep{Zakhozhay13}:
\begin{eqnarray}
& &\log\tau_{\rm MS}({\rm yr})\approx9.96-3.32\log m+0.63(\log m)^2\nonumber\\
& &+0.19(\log m)^3-0.057(\log m)^4.
\label{mslifetime}
\end{eqnarray}
where $\tau_{\rm MS}$ is given in units of years.
Inserting Equations (\ref{pdimf}) and (\ref{mslifetime}) into Equation (\ref{sner}) gives an estimated value of $A_{\rm OB}\approx 4.4-8.8\times10^6 ~{\rm stars}$.
Adopting this normalization, the total number of massive stars in the Galactic disk becomes $\int_{10{\rm M_{\odot}}}^{150{\rm M_{\odot}}}N(m)dm\approx(0.93-1.85)\times10^5$.
This is fairly consistent with the results of \citet{Reed05}, who predicted that the number of stars more massive than $10~\smass$ inside
the solar circle is  $N(>10\smass)\approx2\times10^5$ and that the Galactic ${\rm SNER} \approx 1 - 2$ per century.
We will use the mass distribution function in Equation (\ref{pdimf}) with $A_{\rm OB}\approx 6.6\times10^6 ~{\rm stars}$ to estimate
the relative contribution of wind mechanical energy from stars of different masses.

\begin{figure*}[t!]
\centering
\includegraphics[trim=2mm 2mm 2mm 2mm, clip, width=150mm]{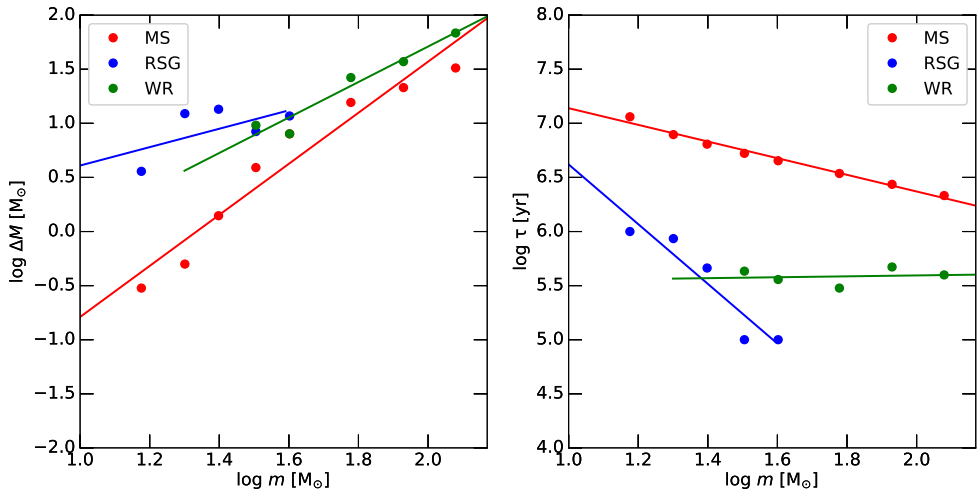}
\vskip 0.0cm
\caption{Left-hand panel: Mass loss, $\Delta M_k(\smass)$, as a function of the initial mass $m$, where $k$ stands for the MS, RSG, and WR stages.
The solid lines show our fitting forms in Equations (\ref{dmms})-(\ref{dmwr}).
Right-hand panel: Lifetime, $\tau_k(\rm yr)$,  as a function of the initial mass $m$ of the MS, RSG, and WR stages.
The solid lines show our fitting forms in Equations (\ref{tms})-(\ref{twr}).
Both $\Delta M_k$ and $\tau_k$ are calculated with the grid of the stellar evolution computation of \citet{ekstrom2012}.
\label{fig:f2}}
\end{figure*}
\subsection{Massive Star Evolution}

Figure \ref{fig:f1} shows the evolution of the stellar mass, $M(t)$, and the mass loss rate, $\Mdot(t)$, for massive stars ($m=15-120~\smass$),
which are adopted from the grid of the stellar evolution computation for {\it nonrotating} stars reported by \citet{ekstrom2012}.
As mentioned in the Introduction, stellar rotation greatly affects stellar evolution, leading to the prediction of different evolutionary tracks
in the Hertzsprung-Russell diagram, and resulting in different mass losses $\Delta M$ and lifetimes $\tau$ at different stages.
Without rotation, for example, stars with $\gtrsim 25~\smass$ experience the WR stage, while stars with $\gtrsim 40~\smass$ do not go through the RSG stage.
Stellar rotation reduces the former mass limit to $\sim 20~\smass$ and the latter mass limit to $\sim 32~\smass$ \citep{georgy2012}.
Thus, the parametrizations for stellar properties such as $\Mdot$ and $\vterm$ adopted in this work should be taken as approximations
with uncertainties of at least a factor of a few.
Figure \ref{fig:f1} demonstrates that the mass loss rate during the MS stage depends strongly on the initial mass with a range of $10^{-8}-10^{-5}~\Msunyr$.
Although $\Mdot(t)$ stays more or less constant during the MS stage, its time variation increases drastically afterwards.

The left-hand panel of Figure \ref{fig:f2} shows the mass loss $\Delta M_k(m)$ during the MS, RSG, and WR stages for nonrotating stars, which are taken from Table 1
of \citet{georgy2013}. This was based on the stellar evolution grid of \citet{ekstrom2012}.
Stars with $25~\smass$, for example, lose about 1.4, 13.5, and 0.4 $\smass$ during the MS, RSG, and WR phases, respectively.
But stars with $120~\smass$ lose up to $100~\smass$, i.e., about 32.5 and 68.4 $\smass$ during the MS and WR phases, respectively.
The solid lines show our fitting forms for $\Delta M_k(m)$:
\begin{eqnarray}
\log \Delta M_{\rm MS}(m) \approx  2.36 \log m - 3.15,\label{dmms}\\
\log \Delta M_{\rm RSG} (m) \approx 0.85 \log m -0.24,\label{dmrsg}\\
\log \Delta M_{\rm WR}(m) \approx 1.64 \log m - 1.57,\label{dmwr}
\end{eqnarray}
where $k$ stands for the MS, RSG, and WR stages.

Moreover, we estimate the lifetime of each stage by identifying the epoch $t$ that corresponds to the mass coordinate at the end of each stage
(e.g.,~$M_{\rm end,MS}= m - \Delta M_{\rm MS}$) in the evolutionary tracks of
 \citet{ekstrom2012}. The right-hand panel of Figure \ref{fig:f2} shows $\tau_{k}(m)$.
The solid lines show our fitting forms for $\tau_{k}(m)$:
\begin{eqnarray}
\log \tau_{\rm MS}(m)\approx -0.77 \log m + 7.91,\label{tms}\\
\log \tau_{\rm RSG}(m)\approx -2.76 \log m + 9.38,\label{trsg}\\
\log \tau_{\rm WR}(m)\approx 0.042 \log m + 5.51,\label{twr}
\end{eqnarray}
where the lifetimes are expressed in unit of years.
For $m\ge 10~\smass$, Equation (\ref{tms}) matches approximately the MS lifetime given in Equation (\ref{mslifetime}), which is based on the stellar evolution calculation by
\citet{Schaller92}.

\subsection{MS and WR Winds}
\label{mswrwind}

Stellar winds from hot massive stars are thought to be driven by the transfer of energy and momentum from the radiation field to the atmospheric gas
through atomic line transitions \citep[e.g.,][for a review]{puls2008}.
So wind parameters such as the mass loss rate and terminal velocity can be estimated
by solving the complex dynamic equations numerically with the line acceleration, which includes radiative
transfer calculations
\citep[e.g.,][]{Vink00,Vink01,Krticka10,Muijres12}.
Just like the star formation process, the theoretical determination of $\Mdot$ and $\vterm$ is a very challenging problem,
because it involves uncertain physics concerning non-LTE processes, opacity, wind clumping, magnetic fields, turbulence, and stellar rotation \citep{Vink2015}.

The mass loss rate, $\Mdot = 4\pi r^2 \rho(r) \vterm$, depends on the density $\rho(r)$ at a radius where the wind has reached its
terminal velocity, $\vterm$.
This terminal velocity can be determined empirically by analyzing the P Cygni profiles of $H_\alpha$ and UV resonance lines,
while $\Mdot$ can be estimated by adopting a wind density model \citep[e.g.,][]{Lamers93, Lamers95,puls2008}.
\citet{Lamers95} measured $\vterm$ of stellar winds from stars of O-F types by analyzing the P Cygni profiles of UV lines.
They found that the observed ratio of $\vterm/\vesc$ changes abruptly from $1.3 $ to $2.6$
at $T_{\rm eff}\approx 2.5\times 10^4$K.
Here, $\vesc=(2GM(1-\Gamma_{\rm e})/R)^{1/2}$ is the photospheric escape velocity corrected for the radiation pressure by electron scattering, parametrized by $\Gamma_{\rm e}$.
This so-called `bi-stability' of winds comes from a shift in the ionization balance of iron (Fe III) that dominates the line acceleration in the
lower part of the wind flow \citep{Vink99}.
Although uncertainties in the empirical values of $\vterm/\vesc$ are estimated to be about 30-40\%,
theoretical predictions indicate much larger variations of this ratio, ranging $1.0\lesssim \vterm/\vesc \lesssim 5.5$ \citep[e.g.,][]{Krticka14,Muijres12}.

\begin{figure*}[t!]
\centering
\includegraphics[trim=2mm 2mm 2mm 2mm, clip, width=160mm]{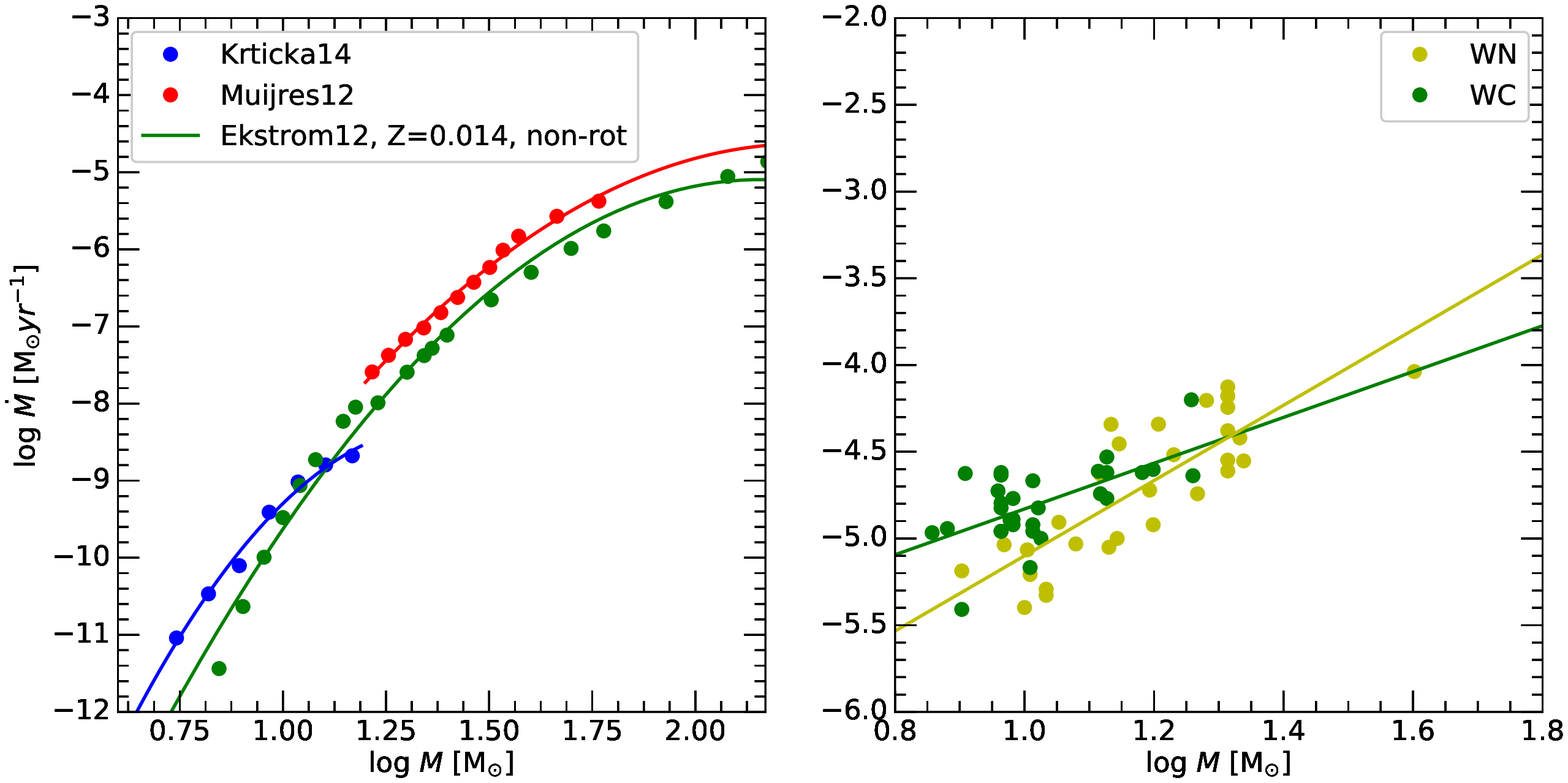}
\caption{Left-hand panel: Mass loss rate $\Mdot$ of MS winds as a function of the MS mass $M_{\rm MS}$.
Theoretical predictions are taken from \citet{ekstrom2012} (green dots),
\citet{Krticka14} (blue dots), and \citet{Muijres12} (red dots).
Estimates of \citet{ekstrom2012} and \citet{Muijres12} were calculated based on the VKL01 recipe.
Right-hand panel: Mass loss rate  $\Mdot$ of WR winds as a function of the WR mass $M_{\rm WR}$.
Observational data for WN (yellow dots) and WC (green dots) stars are taken from \citet{Nugis00}.
The solid lines show our fitting formulas in Equations (\ref{mdms})-(\ref{mdwc}).
\label{fig:f3}}
\end{figure*}

In general, these wind parameters depend on stellar properties such as the stellar mass, $M$, luminosity, $L$, effective temperature, $T_{\rm eff}$, and
metallicity, $Z$, at a given stage.
For example, the theoretical prescription for $\Mdot(L,M,T_{\rm eff},\vterm/\vesc,Z)$ presented by \citet{Vink01} (VKL01) is widely adopted in stellar evolution calculations \citep[e.g.,][]{ekstrom2012}.
Note that $M$ is usually smaller than the initial mass, $m$, due to mass loss.
Moreover, the theoretically predicted values of both $\Mdot$ and $\vterm$ are expected to change significantly with time during the RSG and WR stages,
while they are relatively constant during the MS stage \citep[e.g.,][]{freyer2003,georgy2012}.
As a result, the wind parameters may not be represented accurately by some simple functions of $M$ only,
and the distribution of their observed values may have substantial variances for a given mass range.
So here we attempt to model `time-averaged' values of $\Mdot(M)$ and $\vterm(M)$ as functions of $M$
in order to obtain the wind mechanical power at a given evolutionary stage.

For MS winds, we extract $\Mdot$ at ZAMS from the stellar evolution grid for non-rotating stars with $Z=0.014$ of \citet{ekstrom2012}, which is
based on the VKL01 recipe.
We also take more recent theoretical estimations for $\Mdot$ given in
Table 2 of \citet{Krticka14} for B-type stars and Table 1 of \citet{Muijres12} for O-type stars.
The left-hand panel of Figure \ref{fig:f3} compares $\Mdot$ for MS stars taken from these three references, where $\Mdot$ is given in units of $\Msunyr$.
For the case of \citet{Muijres12}, we choose the results based on the VKL01 recipe, which are larger by a factor of 2-3 than
the values based on their different wind models.
The predictions from both \citet{Krticka14} and \citet{Muijres12} are larger by a factor of two or so than those of \citet{ekstrom2012}.
This demonstrates the levels of uncertainties in the theoretical predictions for the wind parameters.
Below we adopt the fitting form for the results of \citet{ekstrom2012} (green solid line).

\begin{figure*}[t!]
\centering
\includegraphics[trim=2mm 2mm 2mm 2mm, clip, width=160mm]{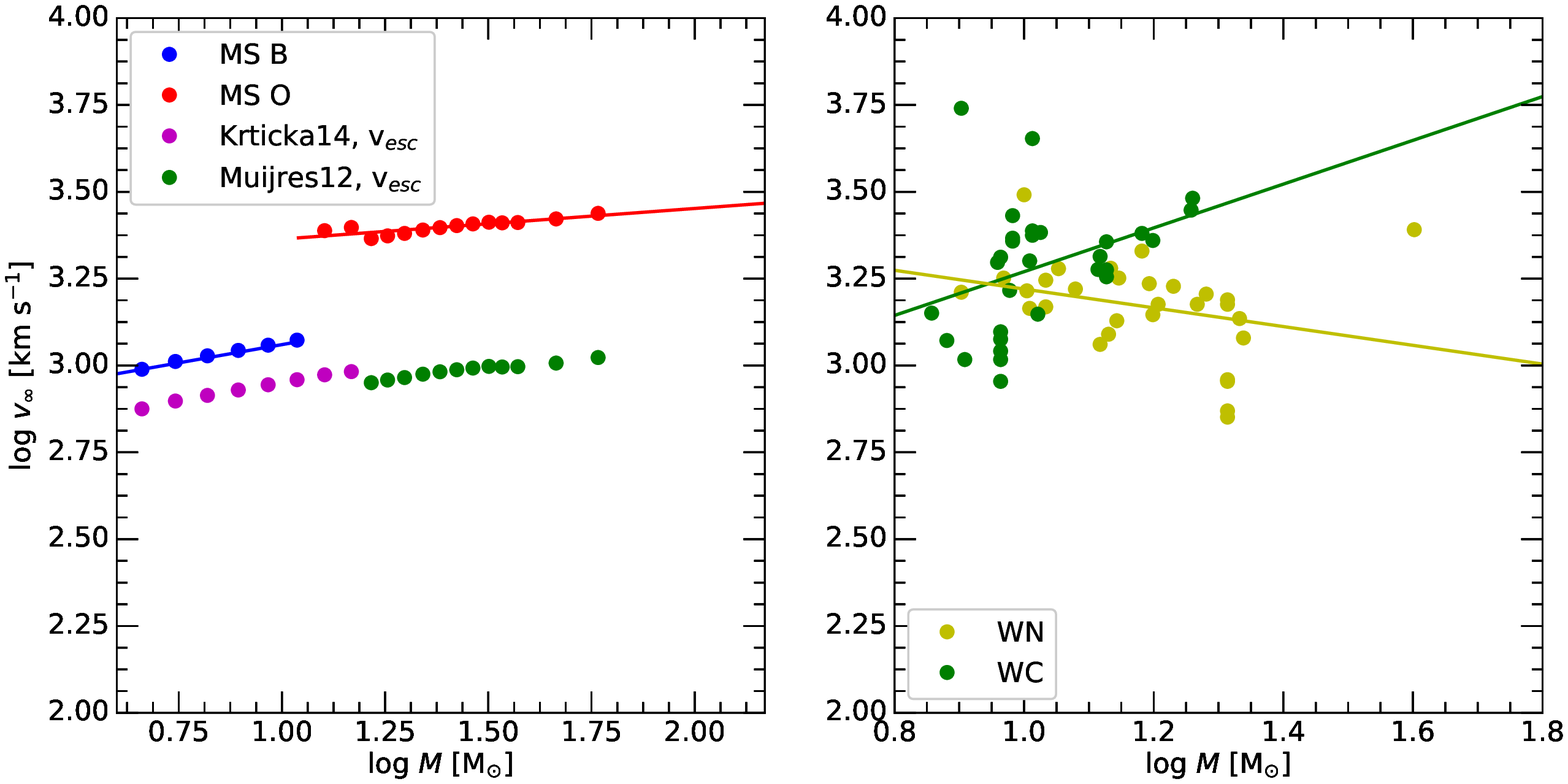}
\caption{Left-hand panel:
Photospheric escape velocity, $\vesc$, and terminal velocity, $\vterm$, for MS stars as a function of the MS mass $M_{\rm MS}$.
Theoretical predictions for $\vesc$ are taken from \citet{Krticka14} (purple dots) and \citet{Muijres12} (green dots),
while $\vterm= 1.3\vesc$ (blue dots) for $M<12~\smass$ and $\vterm= 2.6\vesc$ (red dots) for $M>12~\smass$.
Right-hand panel: Terminal velocity $\vterm$ of WR winds as a function of the WR mass $M_{\rm WR}$.
Observational data for WN (blue dots) and WC (green dots) stars are taken from \citet{Nugis00}.
The solid lines show our fitting formulas in Equations (\ref{vbms})-(\ref{vwc}).
\label{fig:f4}}
\end{figure*}

The theoretical predictions for $\vterm$ given in \citet{Krticka14} and \citet{Muijres12}
tend to have somewhat large variations in the ratio of $\vterm/\vesc$.
So we adopt their estimates of $\vesc$ and use the empirical relations suggested by \citet{Lamers95}, that is,
 $\vterm = 2.6\vesc$ for stars earlier than B1 and $\vterm = 1.3 \vesc$ for stars later than B1.
Here, the spectral type B1 corresponds to $T_{\rm eff}\approx 2.5\times 10^4$~K and $M\approx 12~\smass$.
The left-hand panel of Figure \ref{fig:f4} shows $\vesc$ for MS stars taken from those two references
and $\vterm$ with the aforementioned bi-stability at $ M\approx 12~\smass$,
where the velocities are given in units of $\kms$.

For WR winds, on the other hand, we take the observational data for $\Mdot$ and $\vterm$
in Table 5 for WN stars and Table 6 for WC stars reported by \citet{Nugis00}.
They are shown in the right-hand panels of Figures \ref{fig:f3} and \ref{fig:f4}.
As mentioned above, the observed data points for the WR stage exhibit significant scatter when plotted as a function of $M$
because of the time-variation of the wind parameters.

The solid lines in Figure \ref{fig:f3} show our polynomial fitting forms for $\Mdot$:
\begin{eqnarray}
\log\dot{M}_{\rm MS}\approx -3.38(\log M)^2+14.59\log M-20.84,\label{mdms}\\
\log\dot{M}_{\rm WN}\approx 2.17\log M-7.27,\label{mdwn}\\
\log\dot{M}_{\rm WC}\approx1.32\log M-6.15,\label{mdwc}
\end{eqnarray}
where $M$ represents the mass at a given stage, i.e., $M_{\rm MS}$ and $M_{\rm WR}$ at the MS and WR stages, respectively.
Similarly, the solid lines in Figure \ref{fig:f4} show our linear fitting forms for $\vterm$:
\begin{eqnarray}
\log v_{\rm \infty,B}\approx0.21\log M+2.85,\label{vbms}\\
\log v_{\rm \infty,O}\approx0.08\log M+3.28,\label{voms}\\
\log v_{\rm \infty,WN}\approx-0.27\log M +3.49,\label{vwn}\\
\log v_{\rm \infty,WC}\approx0.63\log M +2.64.\label{vwc}
\end{eqnarray}
The bi-stability transition of $\vterm/\vesc$ occurs at B1 stars with $\sim 12~\smass$. But, for simplicity's sake,
we refer stars earlier than B1 as ``O-type" MS stars and stars later than B1 as ``B-type" MS stars hereafter.
The mass loss rate varies over a wide range as a function of $M$: $\Mdot \sim 10^{-10}-10^{-5}~\Msunyr$ in the MS stage
and $ \sim 10^{-5.5}-10^{-4.0}~\Msunyr$ in the WR stage.
On the other hand, the terminal velocity depends only weakly on $M$: $\vterm \sim 10^{3.0}-10^{3.4}~\kms$ during the MS stage
and $\vterm \sim 10^{3.0}-10^{3.8}~\kms$ during the WR stage.
Although both $\Mdot$ and $\vterm$ for the WR stage show somewhat significant scatters by a factor of up to $3-6$ in Figures \ref{fig:f3} and \ref{fig:f4},
we adopt the simple fitting forms.

\subsection{RSG Winds}
\label{rsgwind}

The basic mechanism driving mass loss from cool RSG stars involves the pulsation of outer layers with ensuing dust condensation
and the acceleration of dust by radiation pressure, which is yet to be fully understood \citep{smith2014}.
The observed values range as $\vterm \sim10-40~\kms$ (with $\vterm/v_{\rm esc}\sim 0.2-0.8$) and $\Mdot \sim 10^{-6}-10^{-4}~\Msunyr$ \citep{dejager1988,jura1990,mauron2011, smith2014}.
Since a quantitative physical model for cool winds is not available yet, here we consider an empirical parametrization due to
\citet{dejager1988}, who constructed an interpolation formula that reproduces observed values of $\Mdot$ for stars of O-M types.
Later \citet{Nieuwenhuijzen1990} published the following slightly adjusted form:
\begin{equation}
\log \Mdot= -7.93 + 1.64 \log L + 0.16 \log M -1.61\log T_{\rm eff},\label{mdrsg2}
\end{equation}
where $L$ and $M$ are given in units of solar values.
Adopting the $M-L$ relation for RSG stars, $M\approx 0.14 L^{0.41}$ \citep{mauron2011}, and taking the average value of $T_{\rm eff}\approx 3750$~K,
Equation (\ref{mdrsg2}) can be approximated as a function of $M$ only,
\begin{equation}
\log \Mdot_{\rm RSG} \approx -10.27 + 4.16 \log M,\label{mdrsg}
\end{equation}
for the RSG mass of $10-32~\smass$.

\citet{mauron2011} constructed a fitting formula for $\vterm$ of the RSG stars observed in the solar neighborhood as
$\vterm \approx 20 \cdot (L/10^5)^{0.35} ~\kms$.
With the $M-L$ relation for RSG stars adopted above, we obtain the following empirical parametrization:
\begin{equation}
v_{\rm \infty,RSG} \approx 1.9 \cdot M^{0.85} ~\kms . \label{vrsg}
\end{equation}

\subsection{Wind Luminosity}
\label{windluminosity}

Adopting the parametrization for $\Mdot$ and $\vterm$ described in Sections \ref{mswrwind} and \ref{rsgwind}, the wind luminosity, $L_{\rm w}$, can be approximated as
\begin{eqnarray}
\log L_{\rm w,B}\approx -3.38( \log M)^2+15.02\log M+20.36,\label{lb}\\
\log L_{\rm w,O}\approx -3.38( \log M)^2+14.77\log M+21.21,\label{lo}\\
\log L_{\rm w,RSG}\approx 5.86\log M+25.79,\label{lrsg}\\
\log L_{\rm w,WN}\approx 1.63 \log M+35.21,\label{lwn}\\
\log L_{\rm w,WC}\approx 2.58\log M +34.63,\label{lwc}
\end{eqnarray}
where $L_{\rm w}$ is given in units of $\rm erg~s^{-1}$, and again $M$ is the mass at a given stage in units of $\smass$.

\begin{figure*}[t!]
\centering
\includegraphics[trim=2mm 2mm 2mm 2mm, clip, width=160mm]{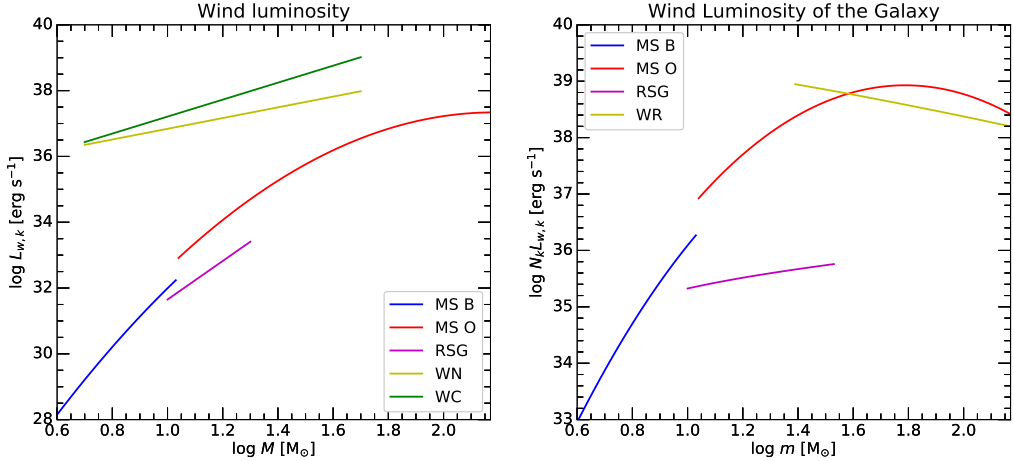}
\vskip 0.0cm
\caption{Left-hand panel: Wind luminosity $L_{{\rm w},k}(M)$ as a function the stellar mass $M$ at different evolutionary stages, where $k$ stands for the MS, RSG, and WR stages.
Right-hand panel: Wind luminosity of the Galaxy $N_k(m)\cdot \langle L_{{\rm w},k}(m)\rangle $ as a function the the initial mass $m$ for different evolutionary stages. Here $N_k(m)$ and $\langle L_{{\rm w},k}(m)\rangle $ are the galaxy-wide IMF and the time-averaged wind luminosity, respectively, for the stars that are born
with the initial mass $m$ and are now in the $k$ phase.
\label{fig:f5}}
\end{figure*}

The left-hand panel of Figure \ref{fig:f5} shows $L_{\rm w}(M)$ at different stages.
In the case of the MS phase, $L_{\rm w,MS}$ increases with increasing $M$ from $10^{32} \ergs$ for $10~\smass$ to $10^{37.5} \ergs$ for $150~\smass$.
The discontinuous change at $ M\approx 12~\smass$ is due to the aforementioned bi-stability of the ratio $\vterm/\vesc$.
The stellar wind power is highest at the WR phase with $L_{\rm w,WR}\approx 10^{36}-10^{39} \ergs $.
Although the mass of observed WR stars ranges $M\approx 5-30~\smass$, they start with the initial mass $m \approx 25-150~\smass$.
As can be seen in Figure 2 of \citet{georgy2013}, $L_{\rm w,WR}(m)>L_{\rm w,MS}(m)$ for a given mass $m$.

\section{Results}
\label{results}

\subsection{Wind Energy Deposition}
\label{windenergy}

Next we estimate how much mechanical energy is deposited by a star with the initial $m$ by integrating the wind luminosity over
the lifetime of a given stage:
\begin{equation}
E_{{\rm w},k}(m) = \int L_{{\rm w},k}(m,t) dt,\label{ek}
\end{equation}
where $k$ represents the MS, RSG, and WR stages.
The wind luminosity formula, $L_{{\rm w},k}(M)$, in Equations (\ref{lb})-(\ref{lwc}) is given as a function of the stellar mass $M(t)$ at a given time.
Since $M(t)$ decreases significantly throughout the lifetime of a massive star, as shown in Figure \ref{fig:f1},
it is not straightforward to relate $M(t)$ with its initial mass $m$, unless we know the mass-loss history of a specific star calculated through stellar-evolution calculations.
For example, there is no {\it a priori} way to find the initial mass $m$ of an observed WR star with $M_{\rm WR}$.

During the MS phase, fortunately, the wind luminosity is almost constant in time for $m\lesssim 32~\smass$,
while it increases slightly by a factor of less than two for $m\gtrsim 32~\smass$ \citep[see Figure 2 of][]{georgy2013}.
As a result, we can assume that the time-averaged luminosity, $\langle L_{\rm w,MS}(m) \rangle$, is similar to $L_{\rm w,MS}(M)$ during the MS stage,
and so the wind energy deposition can be approximated by
\begin{equation}
E_{\rm w,MS}(m)\approx \langle L_{\rm w,MS}(m) \rangle \tau_{\rm MS}(m) \approx L_{\rm w,MS}(M) \tau_{\rm MS}(m),\label{eapprox}
\end{equation}
where $\tau_{\rm MS}(m)$ is the lifetime for the MS phase in Equation (\ref{tms}).

\begin{figure*}[t!]
\centering
\includegraphics[trim=2mm 3mm 2mm 2mm, clip, width=160mm]{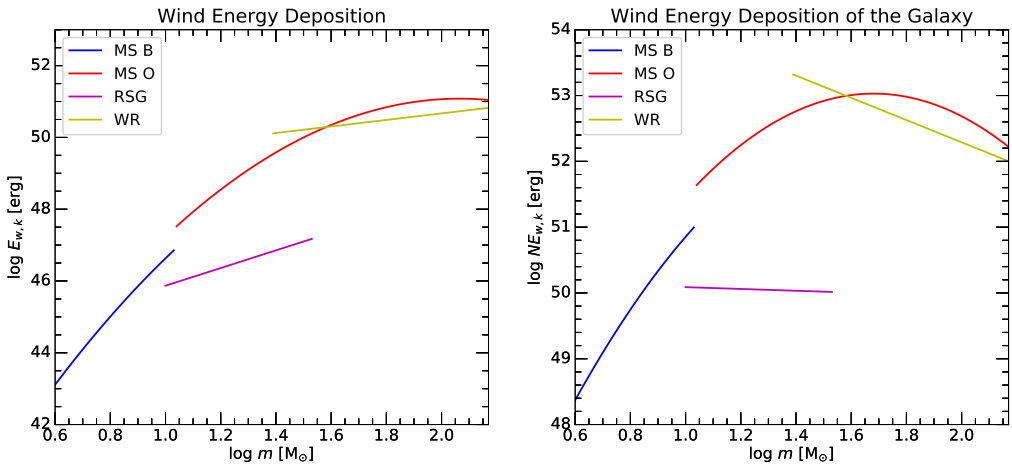}
\caption{
Left-hand panel: Wind energy deposition, $E_{{\rm w},k}(m)= \int L_{{\rm w},k}dt $ (in units of ergs), as a function the initial mass $m$ ($\smass$)
at different stages, where $k$ stands for the MS, RSG, and WR stages.
Right-hand panel: Wind energy deposition of the Galaxy $N(m)\cdot \langle E_{{\rm w},k}(m)\rangle $ ($\rm ergs$) as a function the initial mass $m$
($\smass$) for different evolutionary stages. Here $N(m)$ is the galaxy-wide IMF.
\label{fig:f6}}
\end{figure*}

During the RSG and WR stages, on the other hand, $M(t)$ and $\Mdot(t)$ change significantly.
Moreover, the wind parameters shown in Figures \ref{fig:f3} and \ref{fig:f4} are estimated using the stellar properties such as
$L$, $M$, and $T_{\rm eff}$ at a given time, instead of $m$.
It is not possible to estimate accurately the initial mass from the observed stellar properties
or to convert $L_{{\rm w},k}(M)$ to $L_{{\rm w},k}(m)$ for the RSG or WR phases.
However, using the fact that the terminal velocity is roughly constant during these two stages \citep{georgy2013}, the wind energy deposition can be
approximated as
\begin{equation}
E_{{\rm w},k}(m)\approx \frac{1}{2}v_{\infty,k}^2(m)\int\dot{M}(m)dt = \frac{1}{2} v_{\infty,k}^2(m) \Delta M_{k}(m),\label{eqew}
\end{equation}
where $\Delta M_{k}(m)$ for the RSG and WR phases is given in Equations (\ref{dmrsg}) and (\ref{dmwr}).

Considering that $\vterm(M)$ depends only weakly on $M$, we translate the relations for $\vterm(M)$ to those for $\vterm(m)$ as follows.
The mass at the beginning of the RSG and WR stages are extracted from the stellar evolution grid of \citet{ekstrom2012}
and fitted by the following forms: $\log M_{\rm RSG,i}\approx 0.79 \log m + 0.25$ and $\log M_{\rm WR,i} \approx 1.35 \log m - 0.81$.
Assuming that $M\approx M_{\rm RSG,i}$ for the RSG phase and $M\approx M_{\rm WR,i}$ for the WR phase and
inserting these relations into Equations (\ref{vwn}), (\ref{vwc}), and (\ref{vrsg}), we can obtain approximate relations for $\vterm(m)$, which
are used to calculate $E_{{\rm w},k}(m)$ in Equation (\ref{eqew}).

The left-hand panel of Figure \ref{fig:f6} shows $E_{{\rm w},k}(m)$, during the MS, RSG, and WR stages.
It shows that the time-integrated energy deposition during the MS phase is greater than that during the WR phase,
although $L_{\rm w, WR}(m)$ is higher than $L_{\rm w,MS}(m)$ for a given $m$.
In other words, the MS winds are less energetic but last much longer than the WR winds.
For O stars with $m\gtrsim 80~\smass$, $E_{\rm w,MS}\approx 10^{51}~{\rm ergs}$, which is comparable to the explosion energy of typical SNe.
On the other hands, the RSG winds are less powerful and contribute the least mechanical energy among the three wind types.
Moreover, being cool, dense, and slow winds with a relatively short lifetime, the termination shock of RSG winds is not
expected to be important for production of GCRs.

We assumed above the time-averaged wind luminosity, $\langle L_{\rm w,MS}(m) \rangle \approx L_{\rm w,MS}(M)$ for the MS phase.
For the RSG and WR phases, on the other hand, we calculate
$\langle L_{{\rm w},k}(m) \rangle \approx E_{{\rm w},k}(m)/\tau_k(m)$ by using the estimation for $E_{{\rm w},k}(m)$.
We will use $\langle L_{{\rm w},k}(m)\rangle$ to estimate the galaxy-wide wind luminosity below.

\subsection{Wind Luminosity of the Galaxy}
\label{windluminosityofthegalaxy}

So far, we have estimated the time-averaged wind luminosity, $\langle L_{{\rm w},k}(m) \rangle$, and energy deposition, $E_{{\rm w},k}(m)$, for a star with initial mass $m$.
We now consider the same quantities for all massive stars in the Galaxy with the mass distribution, $N(m)$, in Equation (\ref{pdimf}).
Note that $N(m)dm$ represents the number of stars in the present-day Galaxy that were born with the initial mass in the range of $[m, m+dm]$.
Then we assume that the fraction of stars that are in the $k$ stage is proportional to the lifetime $\tau_{k}(m)$ of each stage as
$f_{k}(m)\approx \tau_{k}(m)/\tau(m)$,
where $\tau_{k}(m)$ is given in Equations (\ref{tms})-(\ref{twr}) and $\tau(m) \approx \tau_{\rm MS}+\tau_{\rm RSG}+\tau_{\rm WR}$.
Then the number of stars in the Galaxy at each stage can be approximated as follows:
\begin{equation}
N_{k}(m)= N(m)f_{k}(m) \approx A_{\rm OB}\cdot m^{-2.6}\cdot\frac{\tau_{k}(m)}{\tau(m)}.\label{numfc}
\end{equation}
This galaxy-wide mass distribution function will be used to estimate the relative contribution of stellar wind luminosity from stars with the initial mass $m$.

The right-hand panel of Figure \ref{fig:f5} shows the galaxy-wide wind luminosity, $N_k(m)\cdot \langle L_{{\rm w},k}(m)\rangle $, from
stars at different phases.
For the MS winds, $L_{{\rm w}, MS}(M)$ increases with $M$ monotonously, but $N_{\rm MS}\cdot \langle L_{\rm w,MS}\rangle$ peaks at $65~\smass$
and then decreases at higher mass due to the power-law mass function.
As mentioned above, $\langle L_{\rm w,RSG}(m)\rangle\approx E_{\rm w,RSG}/\tau_{\rm RSG}$ for the RSG stage,
and $\langle L_{\rm w,WR}(m)\rangle\approx E_{\rm w,WR}/\tau_{\rm WR}$ for the WR stage.
For $m\gtrsim 40~\smass$, the galaxy-wide wind luminosity of the Galaxy due to O-type MS stars is higher than that due to WR stars,
although $\langle L_{\rm w,WR}(m)\rangle$ is higher than $\langle L_{\rm w,MS}(m)\rangle$.
This is because the fraction of MS stars, $f_{\rm MS}(m)$, is much larger than that of WR stars, $f_{\rm WR}(m)$.
The contributions from B-type MS stars and RSG stars to the galaxy-wide wind luminosity are relatively unimportant.

On the other hand, the right-hand panel of Figure \ref{fig:f6} shows the galaxy-wide wind energy deposition, $N(m)\cdot \cdot E_{{\rm w},k}(m)$, which represents
the time-integrated wind mechanical energy deposited during different stages from stars that are born with the initial mass $m$.
Due to the power-law mass distribution, both $N(m)\cdot E_{\rm w,RSG}$ and $N(m)\cdot E_{\rm w,WR}$ decrease with $m$.
Stars with $m\approx 25-60~\smass$ during the MS and WR stages contribute the most wind energy to the ISM.
Again, the contributions from B-type MS stars and RSG stars are negligible.

The total wind luminosity emitted by all massive star in the present-day Galaxy is calculated as follows:
\begin{eqnarray}
{\mathcal L}_{\rm w}=\int_{10\smass}^{150\smass}N_{\rm MS}(m)L_{\rm w,MS}(m)dm\nonumber\\
+\int_{10\smass}^{40\smass}N_{\rm RSG}(m)L_{\rm w,RSG}(m)dm\nonumber\\
+\int_{25\smass}^{150\smass}N_{\rm WR}(m)L_{\rm w,WR}(m)dm.\label{ltotal}
\end{eqnarray}
We find ${\mathcal L}_{\rm w}\approx 1.1\times 10^{41} \ergs$ with the various phases contributing as:
${\mathcal L}_{\rm w,MS_B}\approx 3.2\times 10^{36} \ergs$,
${\mathcal L}_{\rm w,MS_O}\approx 7.3\times 10^{40} \ergs$,
${\mathcal L}_{\rm w,RSG}\approx 7.5\times 10^{36} \ergs$,
and ${\mathcal L}_{\rm w,WR}\approx 4.1\times 10^{40} \ergs$.
So O-type MS stars contribute the most wind luminosity to ${\mathcal L}_{\rm w}$, which
is about 1/4 of the SN luminosity of ${\mathcal L}_{\rm SN} \approx 4.8\times 10^{41} \ergs$.
The galaxy-wide wind luminosity from WR stars is about 10\% of ${\mathcal L}_{\rm SN}$, while
the contributions from B-type MS winds and RSG winds are insignificant.

Since ${\mathcal L}_{\rm w}$ is somewhat smaller than ${\mathcal L}_{\rm SN}$, the relative importance between the two processes in generating GCRs could be
controlled by the CR acceleration efficiencies at shocks associated with stellar winds and SNRs.
As discussed in the Introduction, about 10 \% of SN explosion energy is expected to be transferred to CRs at strong SNR shocks.
However, $\sim 1-10$ \% of the wind mechanical energy might be transferred to CRs at wind termination shocks,
because wind bubbles have complex and turbulent structures that include unstable contact surfaces and multiple shells.
Similar DSA efficiencies are expected also for shocks in PACWBs \citep{debecker2013} and bow-shocks of runaway stars \citep{delvalle2012}.
The results of our study therefore confirm that SNRs are indeed the primary sources of GCRs.

\section{Summary}
\label{summary}

Massive stars are born mainly in the Galactic disk and strongly influence the surrounding ISM through photoionization, stellar winds, and SN explosions.
Mass loss through stellar winds is one of several key processes that govern the evolution of massive stars, but remains to be fully elucidated.
In particular, wind parameters such as mass loss rate and terminal velocity have not been determined accurately because of complex physics involved in
the wind dynamics, such as non-LTE processes, wind clumping, radiative transfer and turbulence among others.

In this study, we attempt to estimate quantitatively the wind mechanical energy deposition from stars more massive than $10~\smass$ in the Galaxy
by adopting the following models:

\begin{enumerate}

\item We assume the Integrated Galactic IMF (IGIMF), $\xi_{\rm IGIMF}(m)\propto m^{-2.6}$.
So the number of stars, $N(m)dm$, formed with the initial mass in the range [$m$, $m+dm$] can be approximated by the power-law form in Equation (\ref{pdimf})
\citep{Kroupa13,Weidner13}.

\item The mass loss rate $\Mdot$ and the wind terminal velocity $\vterm$ can be expressed as functions of the stellar mass $M$ as described in Sections \ref{mswrwind} and \ref{rsgwind} \citep{Vink00, Vink01, Krticka14,Muijres12,Nugis00,mauron2011}. Then the wind luminosity $L_{\rm w}= (1/2)\Mdot \vterm^2$ can be estimated.

\item Assuming that the number of stars in different evolutionary stages is proportional to the lifetimes of each stage, $N_k(m)= N(m)\tau_k/\tau$,
we estimate the contribution of galaxy-wide wind mechanical luminosity from stars at the MS, RSG and WR phases.

\item We use the stellar evolution grid for {\it nonrotating} stars presented by \citet{ekstrom2012} to relate the stellar mass $M(t)$ at a given time
with its initial mass $m$.

\end{enumerate}

Our parametrizations for the wind parameters such as $\Mdot$, $\vterm$, $L_{\rm w}$, and $E_{\rm w}$ should be taken as approximations
with substantial uncertainties, since theoretical models for the massive star evolution and the wind dynamics are not yet fully understood.
With these caveats, we attempt to evaluate the relative importance of stellar winds at different stages.
The main results of this study can be summarized as follows:
\begin{enumerate}

\item The wind luminosity $L_{{\rm w},k}(m)$ at different stages increases with increasing initial mass $m$ of stars.
For a given star, the wind luminosity is strongest during the WR stage with $L_{\rm w,WR}\approx 10^{36}-10^{39} \ergs $(see Figure \ref{fig:f5}).

\item The time-integrated wind energy deposition, $E_{{\rm w},k}(m)$, increases with increasing $m$.
For stars with $m\gtrsim40~\smass$, O-type MS winds with $E_{\rm w, MS} \approx 10^{50}-10^{51}$~ergs
provide the greater energy than WR winds, because the MS lifetime is longer than the WR lifetime.

\item  The galaxy-wide wind mechanical luminosity from stars at the MS, RSG and WR phases is estimated to be
${\mathcal L}_{\rm w,MS_B}\approx 3.2\times 10^{36} \ergs$,
${\mathcal L}_{\rm w,MS_O}\approx 7.3\times 10^{40} \ergs$,
${\mathcal L}_{\rm w,RSG}\approx 7.5\times 10^{36} \ergs$,
and ${\mathcal L}_{\rm w,WR}\approx 4.1\times 10^{40} \ergs$, respectively.
So O-type MS winds provide the greatest amount of wind mechanical power.

\item
The galaxy-wide wind luminosity ${\mathcal L}_{\rm w}\approx 1.1\times 10^{41} \ergs$ is about 1/4 of the SN luminosity
${\mathcal L}_{\rm SN} \approx 4.8\times 10^{41} \ergs$, based on 1.5 SN explosions per century in the Galaxy.

\item It is well established that about 10\% of SN explosion energy can be transferred to CRs via strong blast waves \citep{caprioli14, caprioli2015}.
On the other hand, the CR conversion efficiency for wind mechanical energy from massive stars,
through termination shocks, PACWBs, and bow-shocks of massive runaways, has not yet been estimated quantitatively \citep[see][]{debecker2013,delvalle2012}.
If we adopt $\sim 1-10 \%$ as a somewhat conservative but educated guess, this study confirms SN explosions as the primary origin of GCRs,
while winds from massive stars in pre-supernova stages can provide a significant and complementary contribution.

\end{enumerate}

\acknowledgments{
We thank the anonymous referee and the Editor, S.-C. Yoon, for constructive comments and suggestions.
H.K. was supported by the Basic Science Research Program of the NRF of Korea through grant 2017R1D1A1A09000567.
D.R. was supported by the NRF of Korea through grants 2016R1A5A1013277 and 2017R1A2A1A05071429.


\end{document}